# Increase of resolution by use of microspheres related to complex Snell's law and phase contrast interferometer

Y. Ben-Aryeh [1]

[1] *Department of Physics, Technion-Israel Institute of Technology, Haifa 32000, Israel*

**The increase of resolution by the use of microspheres is related to the use of evanescent waves, satisfying complex Snell's law $n_1 \sin\theta_I = n_2 \sin\theta_T$ where $\sin\theta_I$ and $\sin\theta_T$ are complex trigonometric functions, related to the incident and refracted angles, while the refractive indices $n_1$ and $n_2$ are real. The evanescent waves are obtained in addition to propagating waves satisfying the ordinary Snell's law. Measurements with high resolutions of phase objects like those of semi-transparent biological tissues, is described by an optical system composed of a combination of the microsphere with interferometer.**

*OCIS codes:* (100.6640) Superresolution: (180.4243): Near-field microscopy: (180.3170): Interference microscopy

## 1. INTRODUCTION

There is an extensive amount of works describing enhancement of resolution by the use of microspheres in various imaging systems [1-18]. In the geometric optics approach the dielectric microsphere behaves as a thin lens as the two principal planes of the microsphere optical system are coinciding at the center of the microsphere [19]. According the Abbe resolution limit the minimal lateral distance $d_{min}$ which can be obtained is given by [19]:

$$d_{min} = \frac{\lambda}{2n \sin\phi} = \frac{\lambda}{2NA} . \qquad (1)$$

Here $\lambda$ is the wavelength, and $NA = n\sin\phi$ is the semi numerical-aperture of the lens. As is well known the resolution can be increased by using "immersed lens" i.e. in which the objective is immersed in liquid with index of refraction $n$ (quite often oil is used).

The enhancement of resolution by the use of microspheres , which is much higher than the Abbe limit, is related to the field of scanning near-field optical microscopy (SNOM) [19]. In this field evanescent waves are produced in which one component of the optical wave vector is imaginary, leading to a decay of the wave in this direction. Other components of the wave vector are increasing according to Helmholtz equation, thus decreasing the "effective" wave length in the corresponding plane, and correspondingly increase the resolution.

In the present work it is shown that for evanescent waves, even in geometric optics approach Snell's law remain valid only if the, respectively, incident and transmittance trigonometric functions $\sin\theta_I$ and $\sin\theta_T$ are considered as complex trigonometric functions. We show that such modified Snell's law is useful in analyzing super resolution effects. The present description by which evanescent waves are converted into propagating waves follows the author previous analyses [20-22] but the relations between enhancement of resolutions by microspheres and complex Snell's law are analyzed only in the present work.

There is a certain interesting aspect about the use of microspheres for obtaining high resolutions which was not largely exploited. The measurements which have been made by the microspheres are usually based on light intensity measurement [1-18], so that all the information about the phases of the object, included in the evanescent waves are lost in the measurement process. Information about the phases of the object are especially important for imaging semi-transparent biological tissues for which one would like to learn about lateral changes in the tissue density. There are different phase contrast methods by which phase object can be measured [19, 23]. For a quantitative analysis of a phase object, one uses interference microscopy, and there are many types of interferometers which can be used. In the present work we show how the microsphere can be combined with interferometer so that the image of a phase object like semi-transparent biological tissues can be obtained.

The present paper is arranged as follows:

In section 2 a general analysis is made showing how the fine structures of planar surface of an object are included in the evanescent waves emitted from such plane. An analytical equation is given which gives the enhancement of resolution which can be obtained relative to that of the Abbe resolution limit. This increase of resolution can be obtained only under the condition that the evanescent waves are converted into propagating waves as analyzed in section 3. The geometry of the microsphere is described. Then we develop the condition under which evanescent waves are converted into propagating waves. This analysis is made by using complex Snell's law which includes complex sin and cos functions. The special system of evanescent waves emitted from phase object like that of semi-transparent biological tissue is used as an example in which the microsphere is operated with phase contrast microscope. The complete system for operating a phase contrast microscope with a microsphere is described in Fig. 2 and analyzed in section 4.

## 2. THR INCREASE OF RESOLUTION RELATIVE TO THE ABBE LIMIT BY THR USE OF EVANESCENT WAVES

We analyze in the present section how the spatial structures of the planar surface of an object are included in the EM waves emitted from this surface. Such planar surfaces can appear in various optical systems for which imaging of the upper planar surface is of interest. As an example of such optical system I refer to imaging of corrugated metallic film treated in [20]. In Figures 1 and 2 we are especially interested in imaging of the upper surface of thin film of a phase object.

Consider a monochromatic EM scalar field in homogeneous medium with a refractive index, $n$ given by

$$V(\vec{r},t) = U(\vec{r})\exp(i\omega t) \ . \tag{2}$$

The space-dependent part $U(\vec{r})$ satisfy the Helmholtz equation

$$\left(\Delta^2 + k^2\right)U(\vec{r}) = 0 \ , \tag{3}$$

$$k = nk_0 \ ; \ k_0 = \omega/c \ . \tag{4}$$

Here c is the velocity of light in vacuum (approximately in air), and $k_0$ is the corresponding wave vector. Let us assume that the planar surface of an object is given by $z = 0$ and that the EM field in this plane is given by the Fourier transform

$$U(x,y) = \int_{-\infty}^{\infty}\int_{-\infty}^{\infty} u(k_x,k_y)\exp\left[-i\left(k_x x + k_y y\right)\right]dk_x dk_y \ . \tag{5}$$

Here $k_x, k_y$, are the lateral spatial coordinates in momentum space. For the planar surface of the object, i.e. at $z = 0$, the spatial structures of the object are included in the distribution of the spatial modes $u(k_x, k_y)$. The EM wave propagating from the planar surface of the object into homogenous medium, in the space $z > 0$ with a refractive index $n_1$ is given by

$$U(x, y, z > 0) = \int_{-\infty}^{\infty} \int_{-\infty}^{\infty} u(k_x, k_y) \exp\left[-i(k_x x + k_y y + k_z z)\right] dk_x dk_y . \tag{6}$$

Substituting Eq. (6) into Eq. (3) we get the following Helmholtz equation:

$$\left[k^2 - (k_x^2 + k_y^2 + k_z^2)\right] U(x, y, z > 0) = 0 . \tag{7}$$

This equation satisfies the boundary conditions given by Eq. (5).

Under the condition $k > k_x^2 + k_y^2$, $k_z$ is real and for such case we get propagating waves emitted from the object plane surface:

$$U(x, y, z > 0) = U(x, y, z = 0) \exp(-ikz) . \tag{8}$$

Under the condition $k < k_x^2 + k_y^2$, $k_z$ becomes imaginary and for such case we get the evanescent waves solution:

$$U(x, y, z > 0) = U(x, y, z = 0) \exp(-\gamma k z) \quad ; \quad \gamma = \sqrt{k_x^2 + k_y^2 - k^2} . \tag{9}$$

For evanescent waves, there is a decay of the wave in the $z$ direction. In order to detect the fine structure which is available in the evanescent waves we need to put detectors very near to the plane from which evanescent EM waves are emitted and the microsphere acts as such detector as analyzed in the present work. The resolution is limited by the lateral component of the wavelength $\lambda_T$ given by

$$\lambda_T = \frac{2\pi}{k_T} = \frac{2\pi}{\sqrt{k_x^2 + k_y^2}} . \tag{10}$$

For "propagating" waves $k_T \leq k$, while for evanescent waves $k_T = \sqrt{\gamma^2 + k^2}$. So we get the enhancement of resolution $F$ by the use of evanescent waves as:

$$F \geq \sqrt{\frac{k^2 + \gamma^2}{k^2}} = \sqrt{1 + \frac{\gamma^2}{n_1^2 k_0^2}} . \tag{11}$$

We assumed here that the index of refraction above the object plane is $n_1$

## 3. CONVERSION OF EVANESCENT WAVES TO PROPAGATING WAVES RELATED TO COMPLEX SNELL LAW

We analyze the mechanism by which evanescent waves incident on a microsphere are converted into propagating waves transmitted through the microsphere. It is well known that evanescent waves do not transfer energy into the evanescence direction [19, 21]. This result follows from the fact that for evanescent waves, there is a phase difference of $\pi/2$ between the electric and magnetic field in the plane perpendicular to decay direction of the evanescent waves. Therefore, the Poynting vector in this direction has zero time average. There might be, however, a flow of energy perpendicular to the evanescence direction in a thin layer with a width which is in the order of wavelength. Therefore in order to capture the evanescent waves we need to put detectors very near to the object plane. The microsphere acts as a "tip detector" since the width of the medium between the object and the microsphere surface is less or in the order of a wave length. I find that in the detection of evanescent waves by the microsphere, the plane which is tangential to the microsphere, at a certain incident point, is tilted relative to the horizontal plane of the object. We need therefore to analyze the reflection and refraction of the evanescent waves, which are incident oblique to the surface of the microsphere. Due to the fact that the geometry of the microsphere is well defined, we can analyze for this system the process of converting evanescent waves to propagating waves quantitatively. Similar effects occur in other systems of SNOM.

Very high resolutions have been made for fluorescence biological systems, and it seems that for such systems there is not any limit for the resolution. In a recent paper I have shown that also in this field one needs to take into account uncertainty relations [24]. One should take into account that in fluorescence spectroscopy one cannot get any information on the phases of the object. So, we find that the operation of a phase contrast measurements with a microsphere can lead to important results which are not obtained by fluorescence microscopy.

As we are interested especially in the use of microsphere for measuring a phase object, we describe in Fig.1 the evanescent waves produced by plane EM waves incident perpendicular to a phase object covered by a thin metallic film. Although we describe this special system the analysis made in the present section for the conversion of evanescent waves into propagating waves is quite general for all microsphere experiments. There is a certain misleading point in the usual use of the ordinary Snell's law for geometric optics description of the microsphere experiments. Such approach seems to give quite good results since the imaging in these experiments is produced mainly by propagating waves and the amount of evanescent waves converted into propagating waves is relatively small. However, the modulation of the propagating waves by the evanescent waves, even if its relative intensity is small, it is quite important for resolution phenomena

As described in Fig. 1 a dielectric microsphere with a radius $R$ and refractive index $n_2$ is located above a thin metallic film at a contact point $O$. The medium between the metallic film and the microsphere is with index of refraction $n_1$. Parallel EM waves transmitted through an object in a direction perpendicular to the thin metallic film may be transmitted both as "propagating waves" and "evanescent waves", but the increase of resolution in microsphere experiments is related to the evanescent waves. The microsphere orientation is symmetric relative to a rotation around the $z$ axis connecting the contact point $O$ with the microsphere center $C$. The width $h$ between the metallic smooth surface and a microsphere surface typical point P is given by

$$h = R(1 - \cos\alpha) \qquad . \tag{12}$$

Here $R\sin\alpha = r$ is the horizontal distance along the $x$ direction from $P$ to the contact point O. We assume that $h$ is of order of $\lambda$ or smaller, so that the evanescent waves have not decayed much before the microsphere. The geometric optics approach for microsphere experiments gives fairly good results under the approximation $R \gg \lambda$. But in order that the evanescent waves will not have a large decay before arriving at the microsphere at point $P$, the distance $h$ between $P$ and the metallic film should be of order $\lambda$ or smaller ,i.e., $h \leq \lambda$. Therefore we use the approximation by which for $P$ we have: $(1-\cos\alpha) \ll 1$. This approximation is fairly good for angles $\alpha$ smaller than $30^0$. We do the analysis for conversion of evanescent waves into propagating waves at the point $P$ on the microsphere surface. At this point waves with $(k_x, k_y)$ wave vectors, satisfying the relation $k_x^2 + k_y^2 > (k_0 n_1)^2$, are arriving in the $z$ direction with decay constant $\gamma(k_x, k_y)$. The EM waves arriving at the microsphere surface include both propagating and evanescent waves. As the enhancement of resolution by the microsphere is related to evanescent waves we limit the analysis to such waves. Since the wave vectors arriving at point $P$ are distributed in x, y plane the analysis becomes 3-dimensional.

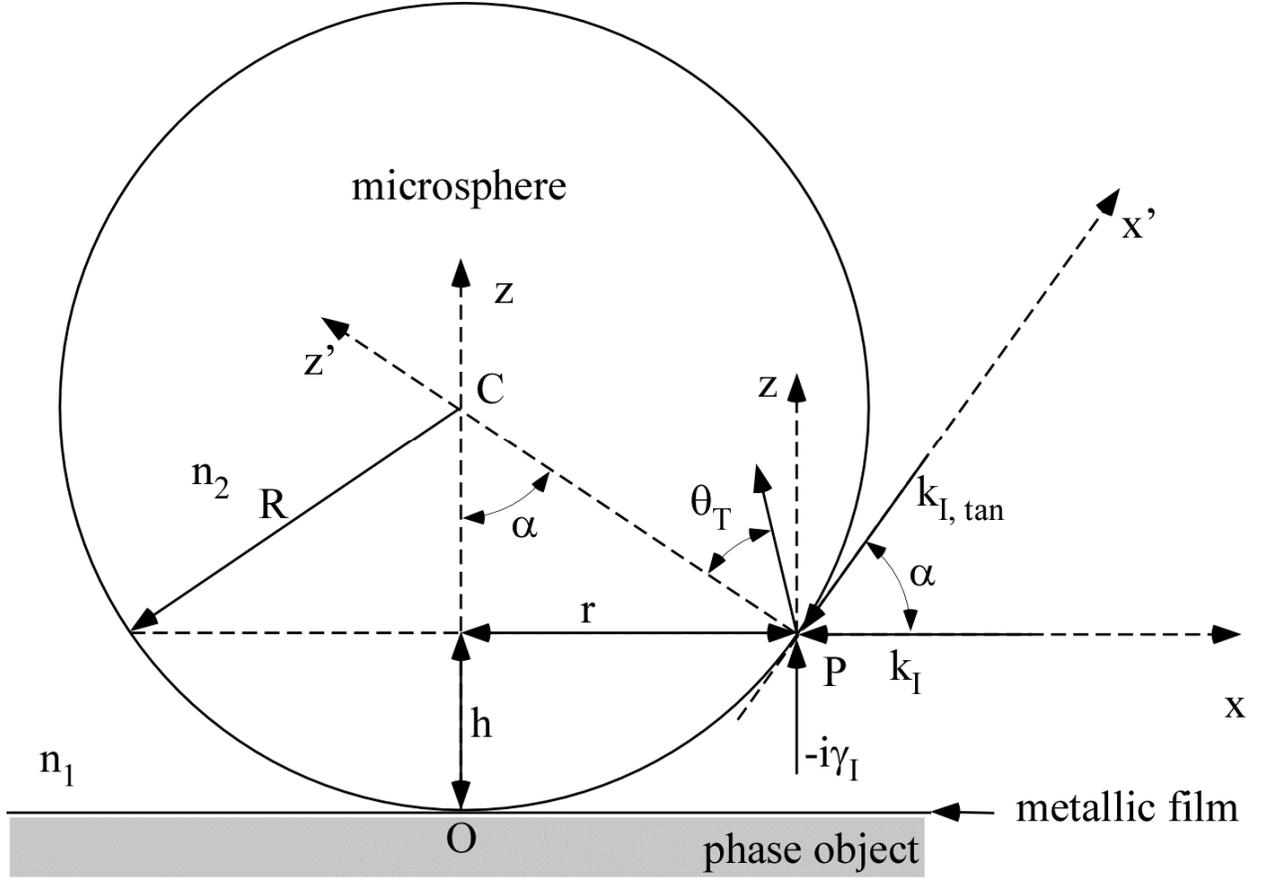

**Fig.1.** A microsphere with refractive index $n_2$ and radius $R$ is located on a metallic film covering a phase object at a contact point $O$. This figure describes the special case where the evanescent waves are incident in the $x, z$ plane, and arriving at the microsphere surface at point $P$ with a wave vector $-k_I \hat{x} - i\gamma_I \hat{z}$ where subscript $I$ denotes incident waves. The component of this wave vector which is parallel to the microsphere surface at point $P$ is given by $k_{I,\tan}$ which is complex and it is preserved during transmission into the microsphere.

. We simplify the analysis by assuming that the EM waves are incident at the point $P$ of the microsphere in the incidence $(x, z)$ plane. Similar analysis can be made if they are incident in the $(y, z)$ plane or any other planar cross section of the microsphere. As described in Fig.1 the wave vector $\vec{k}$ of the evanescent wave emitted from the metallic surface is given by

$$\vec{k} = -k_I \hat{x} - i\gamma_I \hat{z}. \tag{13}$$

Here the subscript $I$ refers to incident wave and $-k_I$, and $-i\gamma_I$ are the components of the complex vector $\vec{k}$ in the $(x,z)$ coordinate system. For the evanescent waves we get

$$k_I^2 = k^2 + \gamma^2 = (n_1 k_0)^2 + \gamma^2 \,. \tag{14}$$

In Fig.1 the $(x',z')$ coordinates system is rotated by angle $\alpha$ relative to the $(x,z)$ coordinate system, where $x'$ and $z'$ coordinates are, respectively, parallel and perpendicular to the microsphere at the $P$ point. The wave vector component $\vec{k}_{I,\tan}$ which is parallel to the microsphere at point P is given by

$$\vec{k}_{I,\tan} = \hat{x}'(-k_I \cos\alpha - i\gamma_I \sin\alpha) \,. \tag{15}$$

We notice that $k_{I,\tan}$ is complex and can be given as

$$k_{I,\tan} = n_1 k_0 \sin\theta_I \quad;\quad \sin\theta_I = (-k_I \cos\alpha - i\gamma_I \sin\alpha)/(n_1 k_0) \,. \tag{16}$$

We find that for incident evanescent wave the incident trigonometric function $\sin\theta_I$ is complex. The complex wave vector component $k_{I,\tan}$ parallel to the microsphere surface at point $P$ does not change at the microsphere boundary surface and is transmitted as

$$k_{T,\tan} = k_{I,\tan} = n_1 k_0 \sin\theta_I = n_2 k_0 \sin\theta_T \,. \tag{17}$$

Here the subscript $T$ refers to the transmitted light. Eq. (17) represents complex Snell's law with complex trigonometric functions, where $\sin\theta_I$ has been defined in Eq. (16).

Evanescent waves incident on the microsphere at point $P$ will be converted to propagating waves entering the microsphere under the condition that the component of the wave vector perpendicular to the microsphere surface, in the microsphere, $k_{T,perp}$ will be real (or equivalently $(k_{T,perp})^2$ will be positive) so that evanescent decay is avoided. We get:

$$(k_{T,perp})^2 = n_2^2 k_0^2 \cos^2\theta_T = n_2^2 k_0^2 (1 - \sin^2\theta_T) = n_2^2 k_0^2 - n_1^2 k_0^2 \sin^2\theta_I \,. \tag{18}$$

On the right hand side of Eq. (18) we used the complex Snell's law given by Eq. (17), where the complex trigonometric functions satisfy the usual equality $\sin^2\theta_T + \cos^2\theta_T = 1$. Substituting the value of $\sin\theta_I$ from Eq. (16) into Eq. (18) we get:

$$\left(k_{T,perp}\right)^2 = n_2^2 k_0^2 - k_I^2 \cos^2\alpha + \gamma^2 \sin^2\alpha - 2ik_I\gamma_I \cos\alpha\sin\alpha. \tag{19}$$

As described in Fig.1, we assumed that the component of the wave vector $\vec{k}_{I,\tan}$ is in the negative $x$ direction. But by taking into account that $\vec{k}_{I,\tan}$ can be both in the positive and negative directions, then the averaged value the imaginary term in (19) vanishes. Substituting Eq. (14) into Eq. (19) and neglecting the imaginary term we get:

$$\left(k_{T,perp}\right)^2 = n_2^2 k_0^2 - n_1^2 k_0^2 \cos^2\alpha - \gamma^2 \cos(2\alpha). \tag{20}$$

For case for which $\left(k_{T,perp}\right)^2 > 0$ i.e. for which $k_{T,perp}$ is real the evanescent waves are converting into propagating waves entering the microsphere without decay. For cases for which: $\left(k_{T,perp}\right)^2 < 0$, i.e. for which $k_{T,perp}$ is imaginary the waves entering the microsphere remain evanescent. Then we find the following fundamental conditions for the light transmitted into the microsphere:

$$\begin{aligned} n_2^2 k_0^2 - n_1^2 k_0^2 \cos^2\alpha - \gamma^2 \cos(2\alpha) > 0 &\rightarrow propagating \\ n_2^2 k_0^2 - n_1^2 k_0^2 \cos^2\alpha - \gamma^2 \cos(2\alpha) < 0 &\rightarrow evanescent \end{aligned}. \tag{21}$$

According to Eq. (21) the conditions for conversion of evanescent waves into propagating waves are improved for a larger index of refraction $n_2$ (as verified by various experiments [3,9,13]). Large values of the decay constant $\gamma$ and small values of $\alpha$ can prevent the conversion of evanescent waves to propagating waves as the term $-\gamma^2 \cos(2\alpha)$ can lead to very high negative values. This factor decreases for large values of $\alpha$ so that the waves remain propagating even for large values of $\gamma$ which increases the enhancement of resolution factor $F$ given by Eq. (11). But on the other hand for large values of $\gamma$ there is a strong decay of the evanescent waves arriving at the microsphere surface.

In order to demonstrate some properties from the present analysis, the evanescent waves amount $\exp(-\gamma h)$, which remained at the distance $h$ before the microsphere surface, is calculated as function of the parameter $(\gamma/n_1 k_0)$ (with a corresponding enhanced resolution factor F) and angle $\alpha$ for the following experimental parameters: $k_0 = 2\pi/4000 = 0.00157(A^0)^{-1}$, $R = 2\mu m = 20000 A^0$, $n_1 = 1.5$, $n_2 = 2$:

**Table 1. Residual amount $\exp(-\gamma h)$ of evanescent waves at the microsphere surface, as function of the angle $\alpha$ and $\gamma/n_1 k_0$ (with a resolution enhancement factor $F$)**

| $\alpha$ | $\gamma/n_1 k_0 = 0.5$ $F = 1.12$ | $\gamma/n_1 k_0 = 1$ $F = 1.41$ | $\gamma/n_1 k_0 = 1.5$ $F = 1.80$ | $\gamma/n_1 k_0 = 2$ $F = 2.24$ |
|---|---|---|---|---|
| $10^0$ | 0.70 | 0.49 | 0.34 x | 0.24 x |
| $20^0$ | 0.24 | 0.0528 | 0.0141 | 0.0035 x |
| $30^0$ | 0.0429 | 0.0018 | $8.0 \cdot 10^{-4}$ | $5.8 \cdot 10^{-6}$ |

In Table 1, $h$ is calculated according to Eq. (12), $\gamma$ is obtained as function of parameter $(\gamma/n_1 k_0)$ and $F$ is calculated according to Eq. (11). One should notice from Table 1 that the residual amount $\exp(-\gamma h)$ of evanescent waves which arrives at the microsphere surface is decreasing for large angles $\alpha$, but for small angles $\alpha$ and large decay constant $\gamma$ the corresponding entries denoted by $x$ represent according to Eq. (21) EM waves which are evanescent waves also in the microsphere. This Table shows that the increase of the resolution, described by the factor $F$ of Eq. (11), is involved with much losses.

There are various diffraction effects which are related to Mie theory and are beyond the geometric optics description. It has been shown in various works that plane waves incident on a microsphere can generate nano jets (see e.g. [25-29]). These nano jets have a subwavelength beam waist and propagate with little divergence for several wavelengths. While various diffraction effects related to the light intensity in the imaging plane and to magnification of the image relative to the object source are important, enhancement of the resolution is related to evanescent waves as analyzed in the present work.

## 4. PHASE CONTRAST INTEROFERMATER WITH A MICROSPHERE

Many objects in microscopy are phase objects which only change the phase of the incident wave without changing the amplitude. Thus, if only the refractive index or thickness of phase object varies across transverse dimensions, then by using an ordinary microscope it will not be able to observe such object. Such an object can be viewed through what is known as phase contrast microscopy [19, 23]. There is, however, interesting work in which microsphere digital holographic approach is presented which realizes phase contrast imaging [30]. The optical system is designed by combining the microsphere optics with the image digital holography. Here we suggest a different optical system which combines the microsphere with an interferometer.

As shown in Fig. 2, plane EM waves are incident perpendicular to the first beam-splitter (BS1), where part of the light is continuing perpendicular to a thin phase object like that of biological tissue covered with metallic film, and part of it is transmitted in horizontal direction and reflected from mirrors M1 and M2. A part of the EM waves emitted from the metallic foil are evanescent and by transmittance through the microsphere they are converted to propagating waves which are recombined in a second beam splitter (BS2) with the reflected waves from mirrors M1 and M2. The recombined beam can be measured by a microscope like that of confocal microscope, not shown in the simple scheme of Fig. 2. A phase shifter (PS) can lead to phase changes between the two combined waves and as analyzed in the present text may lead to phase object measurement.

We follow here a certain approach [31] based on analysis made in the book "Fourier Optics" by J. W. Goodman [32]. Assuming that the image is produced by the object optical field $U_o(x_o, y_o)$ and by the optical field $U_i(x_i, y_i)$ in the image plane then the relation between these two optical fields can be described as

$$U_i(x_i, y_i) = \iint h(x_i, y_i; x_o, y_0) U_0(x_0, y_0) dx_0 dy_0 \quad . \tag{22}$$

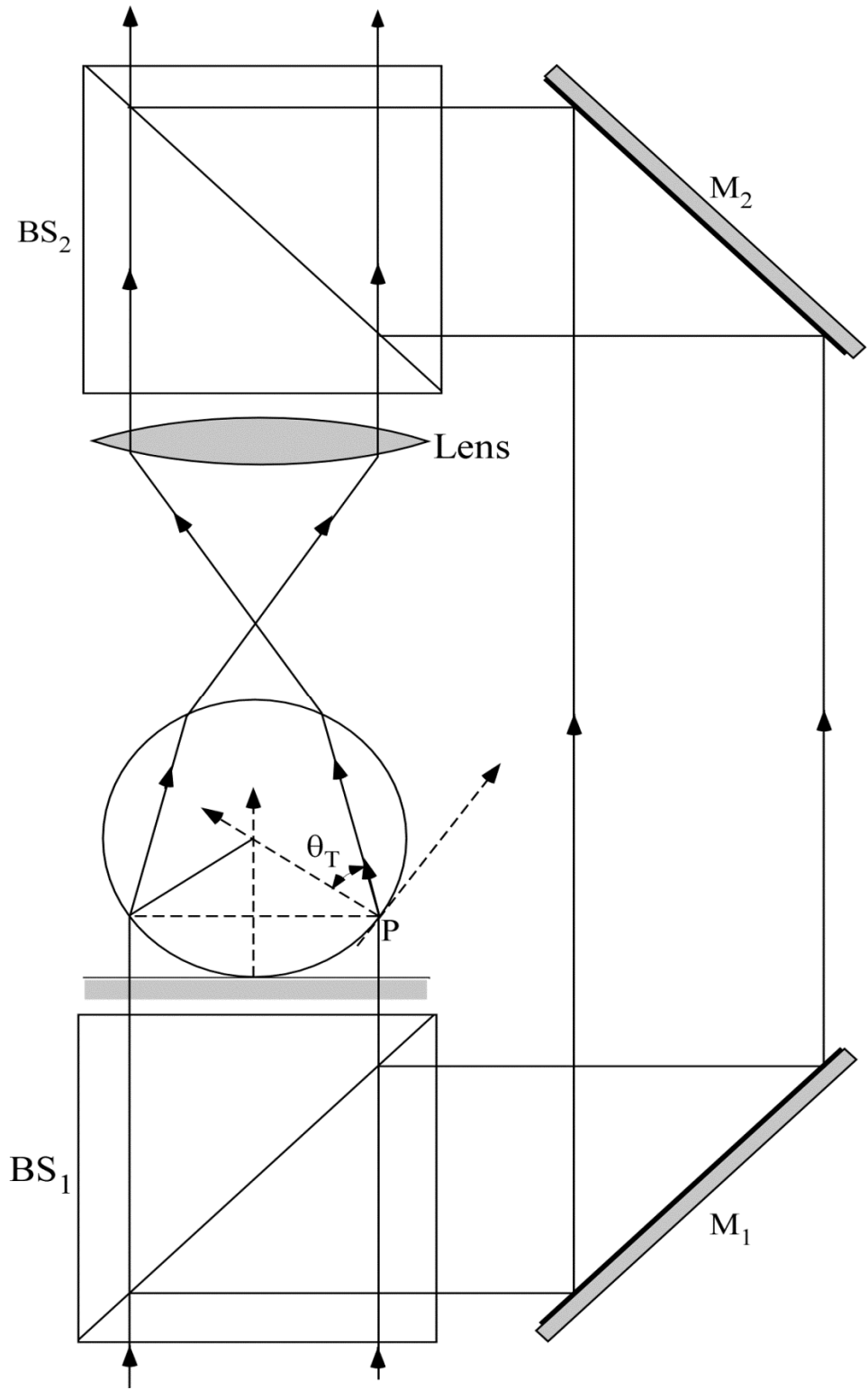

**Fig.2.** A schematic description of the optical system combining a microsphere with an interferometer.

Here $h(x_i, y_i; x_o, y_0)$ is the point spread function [19], and Eq. (22) describes a convolution between the object field and the point-spread function.

In the simple geometric approach in which the microsphere behaves as a thin lens [19] the image plane and the object plane form object-image relation

$$\frac{1}{d_i} + \frac{1}{d_o} = \frac{1}{f} \quad . \tag{23}$$

For the microsphere $f$ is given by

$$f = \frac{(n_2/n_1) R}{2[(n_2/n_1) - 1]} \quad . \tag{24}$$

In most experiments we have $(n_2/n_1) < 2$, and then the light is focused outside the microsphere lens. For the above geometric optics approach a very simple approximation has been given to the spread function (see [31], Eq. (12)):

$$h(x_i, y_i; x_o, y_0) = C \exp\left[\frac{i\pi}{\lambda d_i}\left(1 + \frac{d_0}{d_i}\right)(x_i^2 + y_i^2)\right] \delta(x_i - Mx_o, y_i - My_o) . \tag{25}$$

Here $C$ is a complex constant and we assumed a perfect imaging system magnified by the factor $M = \frac{d_i}{d_o}$, for which a point of coordinates $(x_o, y_o)$ in the object plane becomes a point with coordinates $(x_i = Mx_0, y_i = My_0)$ in the image plane. In this approximation the image field is a magnified replica of the object field multiplied by a paraboloid phase term. In [31] a digital method has been suggested for correction of the phase aberration, and holographic microscopy has been used for phase contrast imaging. I find that under the condition $x_i^2 + y_i^2 \ll \lambda d_i$ in the microsphere system the phase aberration may be neglected in first order approximation but the diffraction effects can be taken into account by estimating the point spread function as:

$$h(x_i, y_i; x_o, y_0) = C\exp\left[-\frac{(x_i - Mx_0)^2}{2\sigma_x}\right]\exp\left[-\frac{(y_i - My_0)^2}{2\sigma_y}\right]\left(\frac{1}{2\pi\sigma_x\sigma_y}\right) = CF(x_i, y_i; x_o, y_o).$$

**(26)**

We defined here the function $F(x_i, y_i; x_o, y_o)$ given in a short notation for the explicit two dimensional Gaussian of (26). Here $\sigma_x$ and $\sigma_y$ represent the width of the point spread function in the $x$ and y coordinates, respectively. For very small values of $\sigma_x$ and $\sigma_y$ Eq. (26) tends to the delta function given in Eq. (25) without the paraboloid phase term. Since we find that a clear imaging has been obtained in the optical experiments with microspheres, we adopt here an empirical approach by which $\sigma_x$ and $\sigma_y$ are considered as experimental parameters.

As described in Fig. 2 the EM beam is divided by BS1 into a part with constant amplitude $U_1$ which is transmitted through the microsphere and another part with constant amplitude $U_2$ by-passing the microsphere. The EM wave with constant amplitude $U_1$ after its transmittance through the thin layer of a phase object can be described as

$$U_1(x_o, y_o) = U_1 \exp[-i\varphi(x_0, y_0)] .$$

**(27)**

We inserted here the phase of the phase object $\varphi(x_0, y_0)$ which is a function of the object $(x_0, y_0)$ coordinates.

Substituting Eq. (27) into Eq. (22) we get:

$$U_1(x_i, y_i) = U_1 \iint h(x_i, y_i; x_o, y_0)\exp[-i\varphi(x_0, y_0)]dx_0 dy_0.$$

**(28)**

Since the usual measurement with microspheres, are for the light intensity the phase terms do not contribute to intensity measurements which are proportional to $|U_1(x_i, y_i)|^2$. In the simple geometric approach the point spread function is given by

$$h(x_i, y_i; x_o, y_0) = C\delta(x_i - Mx_o, y_i - My_o).$$

**(29)**

Substituting Eq. (29) into Eq. (28) we get

$$U_1(x_i, y_i) = CU_1 \exp[-i\varphi(Mx_0, My_0)]. \tag{30}$$

The simple geometric approach leading to Eq. (30) can be changed by using the point spread function of (26) which takes into account empirically diffraction effects. Then we get:

$$U_1(x_i, y_i) + CU_1 \iint F(x_i, y_i; x_o, y_o) \exp[-i\varphi(x_0, y_0)] dx_o dy_o. \tag{31}$$

The second beam after reflection from mirrors 1 and 2 and transmittance through a phase shifter (PS) is given by $U_2 \exp(-i\alpha)$. Here $\alpha$ is a certain phase shift between the two beams which can be changed by a phase shifter (PS). The interference between $U_2 \exp(-i\alpha)$ and $U_1(x_i, y_i)$ gives the EM field:

$$U_{PC}(x_i, y_i) = U_1(x_i, y_i) + U_2 \exp(-i\alpha). \tag{32}$$

The measured light intensity becomes proportional to

$$|U_{PC}(x_i, y_i)|^2 = |U_1(x_i, y_i)|^2 + |U_2|^2 + [U_1(x_i, y_i)|U_2|\exp(-i\alpha) + C.C.]. \tag{33}$$

Here $C.C.$ denotes the complex conjugate of the first product appearing in the square brackets of Eq. (33). In the simple geometric approach we substitute Eq. (30) into Eq. (33) and then we get:

$$|U_{PC}(x_i, y_i)|^2 = |CU_1|^2 + |U_2|^2 + [CU_1 \exp[-i\varphi(Mx_0, My_0)]|U_2|\exp(-i\alpha) + C.C.]. \tag{34}$$

By assuming $CU_1 = |CU_1|\exp(i\theta)$ and choosing the phase $\alpha$ so that $\theta - \alpha = \pi$ we get:

$$|U_{PC}(x_i, y_i)|^2 = |CU_1|^2 + |U_2|^2 - 2|CU_1||U_2|\cos[\varphi(Mx_0, My_0)]. \tag{35}$$

Eq. (35) demonstrates the conversion of the phases of the phase object to light intensities. Although various possible measurements of phase objects have been described [19,23] the special point here is that by the combination of the interferometer with microspheres, which converts evanescent waves to propagating waves, fine structures of phases can be measured. One should take into account that the analysis made in the previous section by which fine structures of the object can be obtained by evanescent waves is valid for any object field including the special phase object $U_1(x_i, y_i)$ given by Eq. (30). The geometric optics approach leading to equation (35) can be changed by taking into account diffraction effects which may exchange the point spread function (29) into the empirical spread function (26). The general effect of such change will be to spread the image by the point spread function. Then by substituting Eq. (31) into Eq. (32) we get:

$$U_{PC}(x_i, y_i) = CU_1 \iint F(x_i, y_i; x_o, y_o) \exp[-i\varphi(x_0, y_0)] dx_o dy_o + U_2 \exp(-i\alpha). \tag{36}$$

The light intensity becomes proportional to

$$\begin{aligned}|U_{PC}(x_i, y_i)|^2 &= |CU_1|^2 + |U_2|^2 + \\ &\left[|CU_1||U_2|\exp[i(\theta-\alpha)]\iint F(x_i, y_i; x_o, y_o)\exp[-i\varphi(x_0, y_0)]dx_o dy_o + C.C.\right]\end{aligned} \tag{37}$$

The general effect obtained by exchanging Eq. (35) to Eq. (37) is to spread the image of the phase object by the point spread function. Each point in the image for $\varphi(Mx_0, My_0)$, obtained by Eq. (35), will be spread according to Eq. (37) over distance changes $(\Delta x, \Delta y)$ which are of order $(\sigma_x, \sigma_y)$,

## 5. Summary

The increase of resolution by the use of microsphere optical systems has been related to SNOM where the microsphere is acting as 'tip' detector converting evanescent waves into propagating waves. The enhancement of resolution relative to that of the Abbe resolution limit has been developed into an analytical equation given by Eq. (11). Using complex Snell law and complex boundary conditions at the microsphere surface, as described in Fig. 1, the condition for converting evanescent waves into propagating waves has been derived in Eq. (21). By using Table 1 it has been shown that the increase of resolution is involved with many losses for the evanescent waves. It is claimed that although the imaging of the object can be obtained mainly by propagating waves the modulation of the image with even relatively small amount of evanescent waves is the main factor for increasing the resolution.

An optical system for measuring phase object like that of semi-transparent biological tissue with high resolution has been suggested and analyzed by using a combination of the optical microsphere system with an interferometer, as described in Fig. 2. The possibility to convert the phases of the object to light intensity measurements has been related to certain empirical equations.

# Acknowledgement

I would like to thank Prof. Steve Lipson for interesting discussions.